\documentclass[aps,prl,twocolumn,showpacs,preprintnumbers,nofootinbib]{revtex4}
\usepackage{amsmath,amssymb,amsfonts}

\newcommand{\bse}{\begin{subequations}}
\newcommand{\ese}{\end{subequations}}
\newcommand{\be}{\begin{equation}}
\newcommand{\ee}{\end{equation}}
\newcommand{\bea}{\begin{eqnarray}}
\newcommand{\eea}{\end{eqnarray}}
\DeclareMathOperator{\dslash}{\partial\!\!\!\! /}
\begin{document}
\preprint{IC/2008/064\cr arXiv:0809.0825}
\title{3 Dimensional ${\cal N}=8$ Supersymmetric Field Theory Revisited:\\ A Superfield Formulation}
\author{Mahdi Torabian}
\email[]{mahdi@ictp.it}\homepage{http://users.ictp.it/~mahdi}
\affiliation{International Centre for Theoretical Physics, Strada Costiera 11,
I-34014, Trieste, Italy}
\date{September 4, 2008}
\begin{abstract}
In [arXiv:0808.1677] 3 dimensional field theories are proposed to on-shell
represent ${\cal N}=8$ SUSY algebra and $SO(7)$ or $SO(4)\times SO(4)$ subgroups of full $SO(8)$ automorphism. They are theories of 8 scalar and 8 spinor matrix fields with Yukawa, quartic and sextic self-interactions. This note is devoted to superfield and off-shell formulation of the theory. Scalar superfields are proposed to represent scalar supermultiplets. A renormalizable  action with quadratic and quartic self-interactions is proposed to govern their dynamics.
\end{abstract}
\maketitle
\section{I. Introduction and Conclusion}
In \cite{T-theory} interacting field theories are proposed to on-shell
represent ${\cal N}=8$ supersymmetry (SUSY) algebra in 3 dimensions. They are theories of 8 scalar and 8 spinor fields with Yukawa, quartic and sextic self-interactions. Dynamical fields are Hermitian $N\times N$ matrices which are valued in a non-associative algebraic structure. As $R$-symmetry of their actions, they are invariant under only $SO(7)$ or $SO(4)\times SO(4)$ subgroup of full $SO(8)$ automorphisms. They can also be promoted to conformal invariant theories by turning off dimensionful mass parameter.

An elegant approach to formulate supersymmetric (SUSic) field theories and describe SUSY representations is the formalism of superfields in superspace. In this note a superfield formulation of the above SUSic field theories is presented. Scalar superfields are presented to incorporate scalar, spinor and auxiliary fields. The action in superspace, contains kinetic term, quadatic mass term and quartic self-interaction of superfields. It can be made scale (and thus conformal) invariant by turning of mass.
Superfields are as well Hermitian $N\times N$ matrices valued in a non-associative algebra.

\section{II. The Superfield Formulation}
In this section superfield formulation in $SO(7)$ and $SO(4)\times SO(4)$ invariant form is presented sequently.
\subsection{i. The $SO(7)$ Invariant Formulation}
To on-shell represent ${\cal N}=8$ SUSY algebra with reduced $SO(7)$ $R$-symmetry, the theory must contain 8 bosonic and 8 fermionic fields. Scalar fields $\phi^i$ and $\phi^8$ are in $\bf 7$ and $\bf 1$ representation of $SO(7)$ respectively. Spinor field $\psi^A_\alpha$ is in $\bf (2_r,8_r)$ representation of $Spin(1,2)\times Spin(7)$. The Grossmann coordinates of superspace $\theta_\alpha^A$ are in the same representation as fermionic fields. The superspace is then parameterized by commutating and anti-commutating coordinates $(x^\mu,\theta_\alpha^A)$ which are Hermitian \cite{Gates} (see also \cite{Petkou,Bandos}). One can construct a differential and integral calculus for Grassmann variables. The derivative with respect to anti-commutating variable is defined as \bea \partial_\alpha^A\theta_\beta^B = \delta_{\alpha\beta}\delta^{AB},\eea where $\partial_\alpha^A=\partial/\partial\theta_\alpha^A$. Hence, the momentum operators $i\partial_\alpha^A$ and $i\partial_\mu$ are Hermitian as well. The
integration over anti-commutating variable is equivalent to differentiation \be \int {\rm d}\theta_\alpha^A\theta_\beta^B =
\delta_{\alpha\beta}\delta^{AB}.\ee Thus, the double integration
becomes \be \int\epsilon^{\alpha\beta}{\rm d}\theta_\alpha^A{\rm
d}\theta_\beta^B\delta_{AB}\epsilon^{\gamma\eta}\theta_\gamma^C\theta_\eta^D\delta_{CD} = \int {\rm d}^2\theta\theta^2 = -1.\ee

The superfield is defined as a function on this superspace. The SUSY algebra is realized on this function in terms of derivatives.
Supersymmetrically invariant derivatives are \be \partial_\mu\quad {\rm and}\quad {\cal D}_\alpha^A = \partial_\alpha^A + i\theta_\beta^A\sigma^\mu_{\beta\alpha}\partial_\mu,\ee where $\partial_\mu=\partial/\partial x^\mu$. 

The simplest representation of SUSY group of transformations is scalar supermultiplet which can be described by a real scalar superfield $\Phi$. It is defined as $\Phi(x^\mu,\theta^A_\alpha) = \Phi'(x'^\mu,\theta'^A_\alpha)$. The transformation law for superfields is defined as
follows \be\label{variation} \delta\Phi = \Phi(x',\theta') - \Phi(x,\theta) =
i\epsilon^{\alpha\beta}\delta_{AB}\epsilon_\alpha^AQ_\beta^B\cdot\Phi, \ee where $\epsilon$ is the parameter and $Q$ is the generator of superspace transformation. It can be represented in differential form as follows \be Q_\alpha^A = \partial_\alpha^A - i\theta_\beta^A\sigma^\mu_{\beta\alpha}\partial_\mu.\ee

To off-shell represent ${\cal N}=8$ SUSY algebra one has to incorporate 8 more auxiliary (non-dynamical) bosonic fields. Moreover, the superfield can be expanded in a terminating Taylor series in $\theta_\alpha^A$. The coefficients are just component fields. Off-shell there exist 8 bosonic and 16 fermionic fields with dimension 1/2 and 1 respectively. In general there exist a huge number of (finitely many) auxiliary fields of higher mass dimensions which are not dynamical. They can be integrated out by their algebraic equations of motion. Specifically, there are 176 auxiliary fields with dimension 3/2. There are a lot more up to dimension 17/2. One can find submultiplets of the supermultiplet which transform into themselves under supersymmetry and are not subject to their equations of motion. One proposes the following superfields \be\begin{split}
\Phi^i(x,\theta) &= \phi^i(x) + \theta^A\gamma^i_{AB}\psi^B(x) \cr &+
\theta^A\gamma^{ijkl}_{AB}\theta^BF^{jkl}(x) + \theta^A\gamma^{ijk}_{AB}\theta^BF^{jk8}(x),
\end{split}\ee \vspace*{-2mm}\be \Phi^8(x,\theta) = \phi^8(x) + \theta^A\psi^A(x) + \theta^A\gamma^{ijk}_{AB}\theta^BF^{ijk}(x),
\ee in $\bf 7$ and $\bf 1$ representations. $\gamma^i$ are gamma matrices of $SO(7)$. Using scalar-vector duality in 3 dimensions, one can easily check that variation of proposed superfields closes on themselves. In the above proposal for the superfield, there exist 8 more auxiliary bosonic fields $\varphi^i$ and $\varphi^8$ defined as\footnote{Or the other way around \bea F^{ijk} &=& t^{ijkl}\varphi^l+t^{ijk}\varphi^8,\\ F^{ij8} &=& t^{ijk}\varphi^k.\eea In this basis superfields are represented as \bea \Phi^i &=& \phi^i+\theta\gamma^i\psi+\varphi^i,\\ \Phi^8 &=& \phi^8+\theta\gamma^8\psi+\varphi^8.\eea} \bea \varphi^i &=& t^{ijkl}F^{jkl}+t^{ijk}F^{jk8},\\ \varphi^8 &=& t^{ijk}F^{ijk},\eea which have mass dimension 3/2, too high a dimension to describe a physical mode. Tensors $t^{ijk}$ and $t^{ijkl}=\epsilon^{ijklmnp}t_{mnp}/3!$ are invariant tensors of $SO(7)$.\footnote{Note that $t^{ijk}\sim\theta^A\gamma^{ijk}_{AB}\theta^B$ and $t^{ijkl}\sim\theta^A\gamma^{ijkl}_{AB}\theta^B$.}

In superspace formalism, SUSY transformations are just coordinate
transformations. Whereas coordinates total derivatives can be ignored, any
superspace integral which does not depend explicitly on the coordinates is
invariant under SUSY transformations. Thus, the action in superspace
is defined as \be\label{definition-action} {\cal S} = \int {\rm d}^3x\int {\rm d}^2 \theta\ {\cal L}[\Phi^i,\Phi^8,{\cal D}_\alpha^A\Phi^i,{\cal D}_\alpha^A\Phi^8].\ee The following expression stands for the most general renormalizable Lagrangian\footnote{For an explicit construction of the action see appendix A.} \be\begin{split}\label{Lagrangian} \!\!{\cal L} = {\rm Tr}&\Big(\big({\cal D}^A \Phi^i\big)^2 + \big({\cal D}^A
\Phi^8\big)^2 \cr &- \Phi^i(m\delta^{ij}\Phi^j+gt^{ijkl}\big[\Phi^j,\Phi^k,\Phi^l,{\cal T}\big]\big) \cr & -\Phi^8\big(m\Phi^8+gt^{ijk}\big[\Phi^i,\Phi^j,\Phi^k,{\cal T}\big]\big)\Big),\end{split}\ee
where $m$ is the mass and $g$ is the coupling constant and $\delta^{ij}$ is the metric of $SO(7)$.\footnote{ Note that to avoid confusion, some numerical coefficients (1/3! and 1/2) are ignored here, they can be easily reimbursed by noting product of anti-symmetric tensors and 4-commutators.}$^,$\footnote{It can be written in more compact form as \be{\cal L} = {\rm Tr}\Big(\big(\Phi^I{\cal D}^2\Phi^I - m\delta^{IJ}\Phi^I\Phi^J - g\tilde t^{IJKL}\Phi^I\big[\Phi^J,\Phi^K,\Phi^L,{\cal T}\big]\Big),\ee where $I=i,8$ and $\tilde t^{ijkl}=t^{ijkl}$ and $\tilde t^{ijk8}=t^{ijk}$. Then, in this basis the Noether charges of supertranslations are \bea {\cal Q}_\alpha^A &\!\!\!=\!\!\!& {\rm Tr}\Big(\Phi^I{\cal D}_\alpha^A\Phi^I\Big), \\ {\cal P}_\mu &\!\!\!=\!\!\!& {\rm Tr}\Big(\partial_\mu\Phi^I\big({\cal D}^2\Phi^I \!\!- \!m\delta^{IJ}\Phi^J -
g\tilde t^{IJKL}[\Phi^J,\Phi^K,\Phi^L,{\cal T}]\big)\Big).\qquad \eea} The 4-commutator is fully anti-symmetric matrix product of 4 ordinary matrices. The fixed matrix ${\cal T}$ is put in the
recipe along the definition of 4-commutator to have a well-behaved one
\cite{TGMT,half-BPS}.

By construction, the action is invariant under superspace group of transformation, bosonic part of which is $SO(1,2)\times SO(7)$ as subgroup of $OSp(8|2)$. It is also invariant under discrete symmetry group ${\cal S}_N$, permutation of $N$ objects for $N\geq 4$ \cite{T-theory}.

The component form of \eqref{Lagrangian} can be achieved by
$\theta$-expansion succeeded by $\theta$-integration
\be\label{component-form}\begin{split} {\cal L} = {\rm
Tr}\Big(&\frac{1}{2}\,\partial_\mu\phi^i\partial_\mu\phi^i + \frac{1}{2}\,\partial_\mu\phi^8\partial_\mu\phi^8 +
i\bar\psi^A\sigma^\mu\partial_\mu\psi^A \cr  +&2^{1/2}\varphi^i\big(2^{-1/2}\varphi^i+m\phi^i +gt^{ijkl}[\phi^i,\phi^j,\phi^k,{\cal T}] \cr &\qquad\qquad\qquad\qquad\quad\ \ + gt^{ijk}[\phi^j,\phi^k,\phi^8,{\cal T}]\big) \cr +&2^{1/2}\varphi^8\big(2^{-1/2}\varphi^8+m\phi^8 \!+gt^{ijk}[\phi^i,\phi^j,\phi^k,{\cal T}]\big) \cr -& \bar\psi^A\big(m\psi^A + g\bar\psi^A\gamma^{ij}_{AB}[\phi^i,\phi^j,\psi^B,{\cal T}] \cr &\qquad\qquad+ g\bar\psi^A\gamma^i_{AB}[\phi^i,\phi^8,\psi^B,{\cal T}]\big)\Big).\end{split}\ee The action \eqref{component-form} is invariant under the following set of component transformations relating bosons and fermions \bea \delta_\epsilon\phi^i &=& \bar\epsilon^A\gamma^i_{AB}\psi^B, \\ \delta_\epsilon\phi^8 &=& \bar\epsilon^A\psi^A, \eea
\be \delta_\epsilon\psi^A = \big(\dslash\phi^i - \varphi^i\big)\gamma^i_{AB}\epsilon^B + \big(\dslash\phi^8 - \varphi^8\big)\epsilon^A,\ee
\vspace*{-5mm}\bea \delta_\epsilon \varphi^i &=& \bar\epsilon^A\gamma^i_{AB}\dslash\psi^B,\\ \delta_\epsilon \varphi^8 &=& \bar\epsilon^A\dslash\psi^A.\eea It can be found from variation of
superfield \eqref{variation} and matching appropriate powers of $\theta$.

Auxiliary fields $F$ can be eliminated using their algebraic equations motion
\bea \varphi^i &=& -2^{-1/2}\big(m\phi^i+ \frac{1}{2}gt^{ijk}[\phi^j,\phi^k,\phi^8,{\cal T}]\cr &&\qquad\qquad\quad\ \ + \frac{1}{3!}gt^{ijkl}[\phi^j,\phi^k,\phi^l,{\cal T}]\big), \\ \varphi^8 &=& -2^{-1/2}\big(m\phi^8+ \frac{1}{3!}gt^{ijk}[\phi^j,\phi^k,\phi^l,{\cal T}]\big).\eea Put it back into the actin, it leads to conventional action for scalars and spinors which was proposed in \cite{T-theory}.

\paragraph{Conformal Symmetry} It is interesting to note that turning off dimensionful parameter $m$, the action would be scale invariant, and it is believed to be invariant under whole $SO(2,3)$ symmetry group of conformal transformations.

\subsection{i. The $SO(4)\times SO(4)$ Invariant Formulation}
To represent ${\cal N}=8$ SUSY algebra, the theory contains 8 bosoninc and 8 fermionic dynamical on-shell fields.
Bosonic fields $\phi^i$ and $\phi^{i'}$ ($i=1,2,3,4$) are in vector representation of either $SO(4)$, say $\bf 4$. Hereafter, unprimed indices refers to first $SO(4)$ and primed indices refers to second $SO(4)$. Fermionic fields $\psi^{\alpha\alpha'}_{\tilde\alpha}$ and $\psi^{\dot\alpha\dot\alpha'}_{\tilde\alpha}$ are in $[(1/2,0),(1/2,0);\bf 2_r]$ and $[(0,1/2),(0,1/2);\bf 2_r]$ representation of $SU(2)_L\times SU(2)_R\times SU(2)_L\times SU(2)_R\times SL(2,\mathbb R)$ (where $\alpha,\dot\alpha=1,2$ are Weyl dotted and un-dotted indices for $SU(2)_L\times SU(2)_R$ subgroup of $SO(4)$ and $\tilde\alpha=1,2$ is Majora index for $Spin(1,2)$, 3 dimensional spacetime Lorentz group.)

The differential and integral calculus can be defined properly. The covariant
fermionic derivatives are defined \bea {\cal D}_\alpha^{\gamma\dot\delta'} &=&
i\partial_\alpha^{\gamma\dot\delta'} +
\theta_\beta^{\gamma\dot\delta'}\sigma_{\beta\alpha}^\mu\partial_\mu, \eea
where $\partial_\alpha^{\gamma\dot\delta'}=
\partial/\partial\theta_\alpha^{\gamma\dot\delta'}$ and
$\partial_\alpha^{\dot\gamma\delta'}=
\partial/\partial\theta_\alpha^{\dot\gamma\delta'}$. The integration is
defined as \be \int {\rm
d}\theta_\alpha^{\gamma\dot\delta}\theta_\beta^{\zeta\dot\eta} =
\delta_{\alpha\beta}\delta^{\gamma\zeta}\delta^{\dot\delta\dot\eta}, \ee and
the double integration as \be \int \epsilon^{\alpha\beta}{\rm
d}\theta_\alpha^{\gamma\dot\zeta}{\rm
d}\theta_\beta^{\delta\dot\eta}\delta_{\gamma\delta}\delta_{\dot\zeta\dot\eta}
\epsilon^{\alpha\beta}\theta_\alpha^{\gamma\dot\zeta}
\theta_\beta^{\delta\dot\eta}\delta_{\gamma\delta}\delta_{\dot\zeta\dot\eta}
=\int {\rm d}^2\theta\theta^2 = -1.\ee There are similar expressions for the
case when upper dotted and un-dotted indices are interchanged.

To off-shell represent ${\cal N}=8$ SUSY algebra and form superfields, one has to include 8 more auxiliary bosonic fields. One considers two sets of superfields, $\Phi^i(x,\theta,\theta)$ and $\Phi^{i'}(x,\theta,\theta)$, which have the following terminating Taylor expansion in Grossmann coordinates     \be\begin{split} \Phi^i &= \phi^i +
\theta_{\alpha}^{\beta\dot\beta'}\sigma^i_{\beta\dot\gamma}
\psi_\alpha^{\dot\gamma\dot\beta'} +
\theta_{\alpha}^{\dot\beta\beta'}\bar\sigma^i_{\dot\beta\gamma}
\psi_\alpha^{\gamma\beta'} \cr &+
\theta_{\alpha}^{\beta\dot\beta'}\sigma^{ijkl}_{\beta\gamma}
\theta_\alpha^{\gamma\dot\beta'}F^{jkl} +
\theta_{\alpha}^{\dot\beta\beta'}\bar\sigma^{ijkl}_{\dot\beta\dot\gamma}
\theta_\alpha^{\dot\gamma\beta'}F^{jkl}\cr &+
\theta_{\alpha}^{\beta\dot\beta'}\sigma^{ij}_{\beta\gamma}\bar\sigma^{k'\!l'}_{\dot\beta'\dot\gamma'}
\theta_\alpha^{\gamma\dot\gamma'}F^{jk'\!l'} +
\theta_{\alpha}^{\dot\beta\beta'}\bar\sigma^{ij}_{\dot\beta\dot\gamma}\sigma^{\!k'\!l'}_{\beta'\gamma'}
\theta_\alpha^{\dot\gamma\gamma'}F^{jk'\!l'},\cr \Phi^{i'} &= \phi^{i'} +
\theta_{\alpha}^{\dot\beta\beta'}\sigma^{i'}_{\beta'\dot\gamma'}
\psi_\alpha^{\dot\beta\dot\gamma'} +
\theta_{\alpha}^{\beta\dot\beta'}\bar\sigma^{i'}_{\dot\beta'\gamma'}
\psi_\alpha^{\beta\gamma'} \cr &+
\theta_{\alpha}^{\dot\beta\beta'}\sigma^{i'\!j'\!k'\!l'}_{\beta'\gamma'}
\theta_\alpha^{\dot\beta\gamma'}F^{j'\!k'\!l'} +
\theta_{\alpha}^{\beta\dot\beta'}\bar\sigma^{i'\!j'\!k'\!l'}_{\dot\beta'\dot\gamma'}
\theta_\alpha^{\beta\dot\gamma'}F^{j'\!k'\!l'} \cr &+
\theta_{\alpha}^{\dot\beta\beta'}\sigma^{i'\!j'}_{\beta'\gamma'}\bar\sigma^{kl}_{\dot\beta\dot\gamma}
\theta_\alpha^{\dot\gamma\gamma'}F^{j'\!kl} +
\theta_{\alpha}^{\beta\dot\beta'}\bar\sigma^{i'\!j'}_{\dot\beta'\dot\gamma'}\sigma^{kl}_{\beta\gamma}
\theta_\alpha^{\gamma\dot\gamma'}F^{j'\!kl},
\end{split}\ee where the coefficients are component fields. Note that Grossmann coordinates $\theta^{\alpha\dot\alpha'}_{\tilde\alpha}$ and $\theta^{\dot\alpha\alpha'}_{\tilde\alpha}$ are in $[(0,1/2),(1/2,0);\bf 2_r]$ and $[(1/2,0),(0,1/2);\bf 2_r]$ representations respectively.

Regarding the action functional of superfields \be {\cal S} = \int {\rm d}^3x\int {\rm d}^2\theta\int {\rm d}^2\theta\ {\cal L}[\Phi,{\cal D}\Phi],\ee one proposes the following as the most general, $SO(1,2)\times SO(4)\times SO(4)$ invariant, renormalizable Lagrangian \be\begin{split} {\cal L} ={\rm Tr}&\Big(\big({\cal D}_\alpha^{\beta\dot\gamma}
\Phi^i\big)^2\!\! + \big({\cal D}_\alpha^{\beta\dot\gamma} \Phi^{i'}\big)^2\!\!
+ \big({\cal D}_\alpha^{\dot\beta\gamma} \Phi^i\big)^2\!\! + \big({\cal
D}_\alpha^{\dot\beta\gamma} \Phi^{i'}\big)^2 \cr &- \Phi^i\big(m\delta^{ij}\Phi^j + g\epsilon^{ijkl}
[\Phi^j,\Phi^k,\Phi^l,{\cal T}]\big) \cr & - \Phi^{i'}\big(m\delta^{i'\!j'}\Phi^{j'} + g\epsilon^{i'\!j'\!k'\!l'}
\Phi^{i'}[\Phi^{j'},\Phi^{k'},\Phi^{l'},{\cal T}]\Big).\end{split}\ee
By construction, SUSY transformations are just translations in superspace.
The component form of the Lagrangian can be derived as
\be\label{mass-deformed-component}\begin{split} {\cal L} = &{\rm
Tr}\Big((\partial_\mu\phi^i)^2 + (\partial_\mu\phi^{i'})^2 +
i\bar\psi^{\gamma\delta}\dslash\psi^{\gamma\delta} + \!i\bar\psi^{\dot\gamma\dot\delta}\dslash\psi^{\dot\gamma\dot\delta}
\cr &+ 2^{1/2}F_{ijk}\big(2^{-1/2}F^{ijk}+m\epsilon^{ijkl}\phi^l+g[\phi^i,\phi^j,\phi^k,{\cal T}]\big) \cr &+ 2^{1/2}F_{i'\!j'\!k'}\!\big(2^{-1/2}F^{i'\!j'\!k'}\!\!\!+
m\epsilon^{i'\!j'\!k'\!l'}\!\phi^{l'}\!\!\!+g[\phi^i,\phi^j,\phi^{k'},{\cal T}]\big) \cr &+ 2^{1/2}F_{ijk'}\big(2^{-1/2}F^{ijk'} + g[\phi^i,\phi^j,\phi^{k'},{\cal T}]\big) \cr & +
2^{1/2}F_{i'\!j'\!k}\big(2^{-1/2}F^{i'\!j'\!k} + g[\phi^{i'},\phi^{j',}\phi^k,{\cal T}]\big)\cr - \bar\psi^{\gamma\eta}&\big(m\psi^{\gamma\eta} \!+g\sigma^{ij}_{\gamma\delta}
[\phi^i,\phi^j,\psi^{\delta\eta},{\cal T}] +
g\sigma^{i'\!j'}_{\eta\delta}
[\phi^{i'}\!,\phi^{j'}\!,\psi^{\gamma\delta},{\cal T}]\big) \cr  -
\bar\psi^{\dot\gamma\dot\eta}&\big(m\bar\psi^{\dot\gamma\dot\eta}+ g\bar\sigma^{ij}_{\dot\gamma\dot\delta}
[\phi^i,\phi^j,\psi^{\dot\delta\dot\eta},{\cal T}] +
g\bar\sigma^{i'\!j'}_{\dot\eta\dot\delta}
[\phi^{i'}\!,\phi^{j'}\!,\psi^{\dot\gamma\dot\delta},{\cal T}] \Big).\end{split}\ee%

The algebraic equation of motion for $F$ are then
\bea 2^{1/2}F^{ijk} &=& -m\epsilon^{ijkl}\phi^l - [\phi^i,\phi^j,\phi^k,{\cal T}], \\ 2^{1/2}F^{i'\!j'\!k'} &=& -m\epsilon^{i'\!j'\!k'\!l'}\phi^{l'} - [\phi^{i'},\phi^{j'},\phi^{k'},{\cal T}],\\
2^{1/2}F^{ijk'} &=& -[\phi^i,\phi^j,\phi^{k'},{\cal T}], \\ 2^{1/2}F^{i'\!j'\!k} &=& -
[\phi^{i'},\phi^{j'},\phi^k,{\cal T}].\eea%
Putting them back into \eqref{mass-deformed-component} one finds the Lagrangian which was proposed in \cite{T-theory}.

\paragraph{Conformal Symmetry} Again it is worth noting that once the dimensionful parameter $m$ is turned off, the action will be scale invariant and thus fully conformal symmetric.
\section{Acknowledgement}
\begin{acknowledgments}
I am thankful to  E. Gava and S. Randjbar-Daemi for insightful discussions.

\end{acknowledgments}
\section{Appendix A. Action Construction}
Explicit formulation of the Lagrangian \eqref{Lagrangian} is presented here (for more details regarding non-associative algebra see \cite{T-theory}). On dimensional grounds one proposes $\big({\cal D}_\alpha^A\Phi^I\big)^2$ to give the kinetic Lagrangian for the scalar multiplet. Interaction terms can be added through a non-linear functional of superfields. In a renormalizable theory it can be at most quartic in superfields. The aim is to look for an action implementing self-interactions among superfields in such a way that preserves full SUSY and $SO(7)$ automorphisms (the case with $SO(4)\times SO(4)$ is straightforward).

Multiplicity of superfields due to extended SUSY enables one to impose additional structures over them to imply how they should be treated collectively. One presumes that besides superspace, superfields are also valued in a 3 dimensional internal (auxiliary) space, {\it i.e.} $\Phi^I(x^\mu,\theta^A_\alpha;\sigma^r)$, local coordinates of which is labeled by $\sigma^r$. It plays the role of group Manifold. From the internal space point of view, superfield and its various superspace derivative are scalar. So is internal space derivative of superfield from superspace stand point $\partial_r\Phi^I$. This derivation does not change the physical dimension of superfield. One proposes the following as renormalizable Lagrangian
\be\begin{split} \!\!{\cal L} = \int{\rm
d}^3&\sigma\Big(\big({\cal D}^A \Phi^i\big)^2 + \big({\cal D}^A \Phi^8\big)^2 \cr & -\Phi^i\big(m\delta^{ij}\Phi^j+ gt^{ijkl}\partial_r\Phi^j\partial_s\Phi^k\partial_t\Phi^l\epsilon^{rst}\big) \cr & -\Phi^8\big(m\Phi^8+ gt^{ijk}\partial_r\Phi^i\partial_s\Phi^j\partial_t\Phi^k\epsilon^{rst}\big)
\Big),\end{split}\ee where $\epsilon^{rst}$ is invariant tensor of internal space isometry group. Using the definition of Nambu
3-bracket \be
\epsilon^{mnp}\partial_m\Phi^I\partial_n\Phi^J\partial_p\Phi^K =
\big\{\Phi^I,\Phi^J,\Phi^K\big\},\ee it can be rewritten more compactly as \be\label{ContinuousLagrangian}\begin{split} \!\!{\cal L} = \int{\rm
d}^3&\sigma\Big(\big({\cal D}^A \Phi^i\big)^2 + \big({\cal D}^A \Phi^8\big)^2 \cr & -\Phi^i\big(m\delta^{ij}\Phi^j+ gt^{ijkl}\{\Phi^j,\Phi^k,\Phi^l\}\big) \cr & -\Phi^8\big(m\Phi^8+ gt^{ijk}\{\Phi^i,\Phi^j,\Phi^k\}\big)
\Big).\end{split}\ee

The existence of Nambu brackets implies that the action is also
invariant under group of 3-dimensional volume-preserving diffeomorphisms of internal space. From the superspace standpoint it is a global symmetry. It is an infinite-dimensional group of transformations and thus the theory has infinite number of degrees of freedom. However, it can be truncated to a finite-dimensional group by a regularization prescription. In fact there is a well-defined recipe for regularization \cite{TGMT}; 
\\ Differentiable functions are mapped into finite size Hermitian matrices \be \Phi(x,\theta,\xi) \rightarrow \Phi(x,\theta)_{rs},\ee for $r,s=1,2,\cdots,N$ where $N$ is the size of matrices. \\ Nambu bracket is promoted to matrix 4-commutators \be
\big\{\Phi^I,\Phi^J,\Phi^K\big\} \rightarrow N[\Phi^I,\Phi^J,\Phi^K,{\cal
T}].\ee The fixed matrix ${\cal T}$ is introduced in the recipe of regularization along the definition of 4-commutator, to have a well-behaved one. It has to be precisely defined \cite{half-BPS}. \\ The volume integration is replaced by trace over matrices \be\int{\rm d}^3\sigma\rightarrow\frac{1}{N}{\rm Tr}.\ee Performing all this on \eqref{ContinuousLagrangian} one is led to \eqref{Lagrangian}. It defines a field theory with finite number of degrees of freedom. In principle the quantization of such a theory is straightforward.

Whereas superfields are now Hermitian $N\times N$ matrices, naively, one would expect that the theory has an $U(N)$ symmetry of global (internal) transformations. From the solution to the equations of motion, it becomes clear that ${\cal T}$ is a traceless matrix which squares to identity matrix. Indeed it is proportional to generalized $Spin(4)$ chirality matrix. One can use $U(N)$ rotations to brings ${\cal T}^5$ in this form. Then, upon precisely defining it, this group is spontaneously broken to ${\cal S}_N$, the group of permutations of $N$ objects \cite{half-BPS}. The Goldstone bosons are just entries of the matrix fields.


\end{document}